# Recent Developments in Cloud Based Systems: State of Art

Mansaf Alam [a] and Kashish Ara Shakil [b]

a, b Department of Computer Science , Jamia Millia Islamia, New Delhi 110025

b Corresponding Author

**Abstract**— Cloud computing is the latest buzzword in the head of techies round the clock these days. The importance and the different applications of cloud computing are overwhelming and thus, it is a topic of huge significance. It provides several astounding features like Multitenancy, on demand service, pay per use etc. This manuscript presents an exhaustive survey on cloud computing technology and attempts to cover most of the developments that have taken place in the field of cloud computing. It discusses about the various available cloud computing platforms, Security in cloud, reference architectures for cloud and storage of data in cloud computing. Furthermore it gives an insight into the recent developments in cloud environment with the help of use cases and the author's perspective about the future of the respective use cases. Finally it concludes by discussing the limitations of data management in cloud and potential research issues in cloud computing that needs to be addressed in future along with a proposed architecture for cloud.

**Index Terms**—Cloud computing, CDBMS, database as a service, cloud platforms

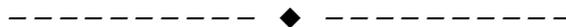

## 1 INTRODUCTION

Cloud computing is a newly emerging technology for the future with it roots based on the rapidly increasing demands on data centers that needs to be catered to. Cloud computing is defined as the use of computing resources to access data over the internet. It is a means or a mechanism to enhance the existing capabilities of Information technology by many folds [120].The terminology cloud comes from the fact that the data is not stored on your desktop or your device but is located far similar to a cloud in literal terms, but despite of it being away its within your reach, you can access it irrespective of your geographical location using a computing device via an internet. Cloud computing is a technology for the future and will change the entire scenario of the IT industry, being a cost efficient approach, with reduced exigency of buying the software or the hardware resources. It is an on demand form of utility computing for those who have access to cloud [38] .Recent web search trends have shown a paradigm shift in peoples interest towards cloud. As per Google search trends [109] there has been an immense increase in people's interest towards cloud computing from 2005 to 2013, also shown by Figure 1.

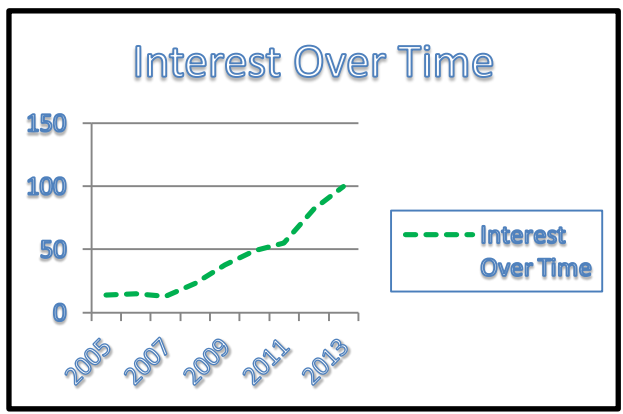

Fig. 1. Google Search Trends for last 8 years

R. Buyya et al.[110] has visioned cloud computing as the 5th utility similar to gas, electricity, water and telephony. They have also described a market oriented architecture for cloud computing.
According to Islam et al. [87] cloud computing succeeds in achieving multilevel virtualization and

b Corresponding Address: Department of Computer Science, Jamia Millia Islamia New Delhi-110025 India

Tel:+91-9899693456

Email address: shakilkashish@yahoo.co.in

abstraction by integrating different varieties of data, its computing storage etc. Virtualization itself can be achieved by means of an encapsulating software layer termed as hypervisor or virtual machine monitor [59].It has made geographical location hurdles for accessing any form of data, service, hardware, software or any other computing resources an extinct phenomenon. In fact, it has played an exquisite role in bringing the most expensive software within the reach of common man through the concept of "pay per use". It's a blessing in disguise for organizations dealing with problems of managing large amount of data along with reduced operational costs. But as every blessing is not perfect in all the forms so is cloud which comes with certain drawbacks in terms of lack of security and privacy and several others which will be discussed later. In fact cloud computing had been rated at number 4th position for the most disruptive technology by Gartner [44].Resource availability is the key for any cloud computing application, and "Pay Per use" being its reward. This paper is an exhaustive survey about the recent developments in the field of cloud computing, the various cloud deployment and service models, cloud platforms available at present, cloud reference architectures etc. The use of cloud computing for storage of data and the use cases of cloud in various fields such as geospatial and air traffic management have also been discussed along with the author's perspective about the future of these use cases. Finally the authors conclude by discussing the limitations of data management in cloud and potential research issues in cloud computing that needs to be addressed in future.

## 1.1 Cloud Deployment Models

Cloud computing comes in four different forms or there are four different models for deployment of a cloud: public, private, hybrid and community [43].

1. **Private cloud:** Private cloud is cloud infrastructure which is managed for a single organization; a private cloud can be owned and managed either internally or externally by an organization. Venturing into a private cloud project requires a significant level and degree of engagement to virtualize the business environment [43]. Sakr et al. [84] suggests that they are widely opposed due their close similarities with the traditional server farms and they do not provide benefits such as no upfront capital costs. For Example Intel, Hewlett Packard (HP) and Microsoft have their own internal private cloud [43].
2. **Public cloud:** Public cloud applications, resources, storage, and other services are made available to the general public by a service provider. These cloud service providers are organizations engaged in selling cloud services like Amazon AWS, Microsoft and Google. These services can be offered either free of cost or may be made available through pay per use. The quality of services offered by the service providers is mentioned in an SLA (Service level agreement) i.e. it is an agreement between a consumer and a cloud service provider. An SLA might include services that will be offered by a service provider in terms of privacy, security, backup procedures [84]. However, public clouds lack fine-grained control over data, network and security settings, which may hamper their effectiveness in different business setups [84]. Some of the examples of public cloud are Amazon Web Services (AWS) and Microsoft Azure [43].
3. **Community Cloud:** In community cloud the cloud infrastructure is shared by the organizations that have common resource requirements (security, jurisdiction and policy), whether managed internally or by a third-party and hosted internally or externally [43]. They take slight advantage of cloud computing in terms of sharing the costs, as the cost is not alone born by a single organization but rather shared by different organizations For Example Google Gov (google apps for government) [43]
4. **Hybrid cloud:** Hybrid cloud is a composition of two or more clouds i.e. it can be a composition of either a private and public or a private and community cloud etc. By utilizing hybrid cloud architecture, organizations and individuals are able to obtain degrees of fault tolerance combined with locally immediate usability without dependency on internet connectivity. Hybrid clouds have few limitations such as the lack of flexibility, security and certainty of in-house client applications. Hybrid cloud ensures the provision and flexibility of in house applications with the fault tolerance and scalability of cloud based services [44].

## 1.2 SERVICE MODELS

Cloud Service providers offer their services through several service delivery models. The various cloud service models are:

1. **Software as a Service**: It provides special purpose software to the end users. They no longer need to install or buy a software product. They can access it directly via an internet for example, Google Apps, Microsoft Office 365. Sales force is Software as a Service provider.

2. **Platform as a service**: Provides a higher level environment through which developers can write customized applications. These include programming language, programming framework. PaaS simplify development of web applications by providing users with facilities such as runtime systems and scalable services etc [67]. For example, Windows azure, Manjrasofts Aneka and Google AppEngine Force.com are some of the platform as service providers.

3. **Infrastructure as a Service**: provides resources required to build an application environment e.g. .servers, storage, network bandwidth and Oracle infrastructure as a service. Amazon EC2 is an IaaS provider. Cloud service model can also be seen by Figure 2[97].

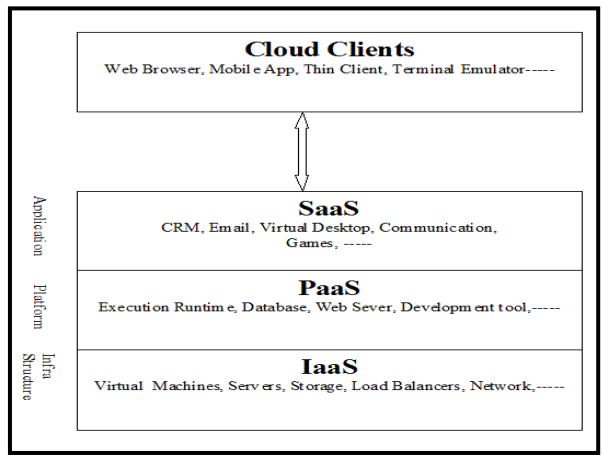

Fig. 2. Layered Cloud Service Model [97]

## 1.3 CHARACTERISTICS OF A CLOUD

Cloud computing has several distinguishing characteristics. These characteristics such as support for virtualization, elasticity, dynamic provisioning of resources etc, offer an attractive option for users to host their applications on cloud. Some of the key characteristics of cloud have been given in table 1.

Table 1.
Characteristics of Cloud

| Characteristics | Description | Advantages | Authors |
|---|---|---|---|
| **Metered Service** | Pricing is based on a pay per use mechanism i.e. computing resources can be used as utilized. | Less initial setup cost, Beneficial for organizations seeking cost cutting, Renting out the already rented resources [82] can further reduce the cloud service costs. | [82] |
| **On-demand self service** | computing resources like server time, database, services etc. can be used by | Ease of use, resources can be added and removed by the users as | [15] |
| | the consumer as per his/her requirement without any human interference | per their requirements | |
| **Size/Population** | Consists of high end servers and computers along with network attached storage. | High end servers and commodity computers available at low costs | [110] |
| **Geographically independent resource access** | Computing resources can be made available across the world irrespective of consumer's location. | Resources can be accessed irrespective of the persons position. | |
| **Elastic** | It provides consumption of resources in an elastic manner i.e. the amount of resources acquired by a particular consumer depends on the usage | Resources can be allocated and reallocated in an elastic manner. | |
| **Infinite storage capacity** | provides its users with an illusion of infinite storage capacity | Users of cloud database can store virtually an infinite amount of data | |
| **Availability** | availability of data is the responsibility of the service providers | Users, no longer have to worry about data backups etc, in cloud it is handled by the service providers. | |
| **Enhanced product development environment** | Allows developers to concentrate on their quality of work they no longer have to worry about availability of servers, recovering from failures, supporting more loads and resources leading to development of quality driven deliverables. | Enhanced quality driven products, availability of better resources, servers etc. | |
| **Service negotiations** | It allows services that are to be provided to be negotiable. The services guaranteed are available in an SLA. | SLA's help in ensuring that certain minimum services must be guaranteed to the users | [110] |

## ORGANIZATION OF PAPER

The remaining portion of this paper has been organized into the following different sections: Section 2 illustrates the motivation for adopting cloud computing i.e. the various use cases of cloud computing such as yahoo cloud service, biomedical information sharing, social cloud, sensor data, geospatial data etc. along with their future, cloud technology being used and author's observations. In section 3 we present about the risks and benefits associated with migrating an organizations IT system to cloud environment. In section 4 we identify a hierarchy of cloud users who are the main users of cloud computing applications. Moving ahead in section 5 a discussion about the cloud reference architectures is done. In section 6 we discuss about the different cloud computing platforms, their key features and advantages of each. Section 7 discusses about the security factor in cloud computing. Furthermore section 8 discusses about the data storage aspect of cloud along with the famous shared disk and shared nothing architectures, database as a service and oracle database 12C. Finally the paper concludes with a discussion on the emerging terminologies and fields in cloud computing section 9 along with open research challenges and a proposed architecture for cloud in section 10 and conclusion and future directions in section 11.

## 2   MOTIVATION FOR ADOPTING CLOUD COMPUTING

The applications of cloud are ubiquitous and immense and have several unique advantages distinct from the ones that are offered by traditional tools and methods. These applications have been recognized in several literatures. In [108] a multi agent based case study utilizing cloud is presented, it leads to reduced costs and increased efficiencies of processes. We have identified numerous applications or use cases of adopting a cloud based approach, these uses cases fall into several areas such as bioinformatics, geosciences, air traffic sensors etc. These use cases clearly demonstrate about the variety of usages and applicability of cloud, and hence the need for moving to a cloud based platform and thus drawing a motivation for adopting cloud.

### 2.1 USE CASES OF CLOUD COMPUTING

Table 2 depicts the various use cases of cloud computing along with the author's observations and perspective about their future using cloud technology. The recent use cases of cloud computing have been enlisted below:

#### 2.1.1   Yahoo cloud service

Yahoo cloud proposed by Cooper et al. [15] provides a horizontal service platform which is shared by various applications which are vertical. There are 3 tiers in yahoo cloud services: core services (server side data management), messaging (e.g. invalidation messages) and edge services (reducing latency and improved delivery).

Hadoop is open source implementations of map reduce parallel processing framework [Cooper et al. 2009][15].It enables developers to concentrate on the development part rather than on details of parallelization. It is an important constituent of yahoo cloud. Hadoop data is stored in hadoop file system which is an open source implementation of Google file system.

#### 2.1.2   Biomedical Information Sharing

Cloud computing can be proved to be very beneficial for biomedical community. As, cloud computing applications can be a blessing in disguise and an answer to several problems relating to information sharing in the field of biomedical sciences. We know information sharing is highly desirable in biomedical science for various scientists to share information with one other. Cloud computing can play several roles in biomedical field such as compute services, archiving data etc [63]. Rosenthal et al.][6] describes how biomedical community including its consortium can take advantage of cloud computing. It also suggests that cloud computing can provide a major improvements in biomedical sciences and should be considered for BMI as it can provide several advantages like less management, highly scalable facilities, better disaster recovery, homogeneity etc. Since this data might contain sensitive information therefore migrating biomedical data to cloud has security concerns. Danilatou and Ioannidis [91] have proposed an architecture to enable secure sharing is proposed based on privacy preserving cryptographic protocols. Data access is governed by security policies.

#### 2.1.3   Geosmart (Social Cloud)

Geosmart is an education based social media in cloud [5].It aims at increasing the intelligence level of people in Indonesia. It is a social cloud in which all the services are accessible by registered Geosmart users and facilitates interaction between users and services. Social cloud has benefits such as low investments, interactive environment and allows individuals to make their contributions for the cause. Members of social cloud can make requests to other members for use of Virtual machines [10] thereby reducing the cost.

### 2.1.4 Managing Sensor data

Huge volumes of sensor data are generated by cloud manufacturing systems and managing them is a task [99].Cloud computing along with relational database and key value store can be used for managing heterogeneous sensor stream data [103]. IOF (Internet of things) is a combination of sensory technology and internet technology [47].

### 2.1.5 Geospatial Data

The geospatial data and its users are growing at an astounding rate these days and the traditional centralized systems cannot support geospatial data storage. Cloud computing along with associated spatial computing is beneficial for highly data intensive and compute intensive geospatial research [53] and can meet the real-time processing demands imposed by geospatial data. Zhong et al. [101] have proposed distributed geospatial data storage and processing framework with characteristics such as cloud based architecture, efficient data placement and map reduce based localized geospatial computing model for parallel processing of data. A cloud model based on spatial data processing can be used as a controller for distribution of spatial data processing along with taking benefits of cloud computing such as supervised and unsupervised classification [29].

### 2.1.6 Air Traffic Management

ATM makes use of elements of cloud computing to achieve a global ATM system and standardizes working procedures for air traffic management [94].In this aircrafts and pilots communicate with each other as cloud participants using a narrowband VHF radio communications. You et al. [77] has proposed a framework for integrating cloud computing and wireless sensor network is proposed, and employs the capability of data processing and services.

### 2.1.7 Big Data Processing

Phenomena of big data involve the act of gathering and processing of enormous data sets [36]. This data set is so large that conventional database systems and software tools have failed to manage it. Big data can be processed and managed by cloud computing in an efficient manner. A detailed survey of future of big data processing by cloud has been presented by Ji et al.[23]. Patel et al. [8] have suggested a solution to big data problem is presented on the basis of experimental result on Hadoop and Map reduce framework. Synchronous parallel processing of big data analysis services can also provide considerable improvement and performance optimization of federated cloud services [49].

Table 2.
Various Use Cases of Cloud

| Use Case | Cloud Technology | Future | Observations | Authors |
|---|---|---|---|---|
| Yahoo cloud service | Hadoop(HDFS) | Quality assurance of yahoo services, scalability and high availability | Will lead to reduced costs and enhanced innovations | [15] |
| Biomedical Information Sharing | Parallel computing using cloud and cloud along with cryptographic techniques | Sharing information amongst biomedical communities, disaster recovery, archiving biological data | Lower costs as compared to a dedicated laboratory | [63],[6],[91] |
| Geosmart(social cloud) | Cloud framework using virtual machines | improving intelligence level of people, low investments, interactive environment | Establishment of virtual organizations | [10] |
| Managing Sensor data | Cloud computing with relational database and key value store, Hadoop | Managing heterogeneous sensor data, | Effective management of large volumes of sensor data | [103],[99],[47] |
| Geospatial Data | cloud model based on spatial data processing, Hadoop and Map Reduce based localized geospatial computing model | Storage and easy management of huge sets of geospatial data, | Efficient sharing and integration of geospatial resources | [53],[101],[29] |
| Air Traffic Management | Software as a Service | Global ATM system, standardized working procedures | Development of Global ATM system, standardized working procedures | [94], [77] |
| Big Data Processing | experimental results based on Hadoop and Map reduce framework | Management of enormous data, extracting useful information from the large data. | Management of infinite sized data sets irrespective of their high volume and variety. | [36], [23],[49] |

## 3 MIGRATION TO CLOUD

There are several benefits and risks associated with migration of an IT system in any enterprise to that of cloud for example to move an IT system in oil and gas industry from an in house data center to Amazon EC2.A solution for rectifying problems in cloud computing such as partitioning of CPU of a single machine into several database appliances on the same machine is given by Aboulnaga et al. [1]. Khajeh-Hosseini et al.[4] have shown a case study of migrating an Enterprise IT system to IAAS. Babar and Chauhan [60] describes the effect of migrating to cloud for Hackystat open source cloud computing software, which provides a guideline for improving software engineering support for developing cloud based systems. Cloud computing business framework [90] provides organizations with capability to design, deploy, and migrate to cloud in a more efficient manner. Four key areas such as classification, organizational sustainability modeling, and portability of services along with any kind of linkages are handled by this model.

Stake holder's impact analysis is done to find out the pros and cons associated with migration of a company to cloud infrastructure .It was carried out through analysis of interview data. It suggests that there are numerous benefits along with risks in migration of a system to cloud. Benefits of migration to cloud are given by table 3 [4].Opportunity to manage income and outgoings is considered to be the key benefit in migrating to the cloud infrastructure .Besides this other benefits have also been identified as some of the key benefits such as opportunity to offer new products and services as all the software's and infrastructures to develop any new product will be made available via cloud, it also removes tedious works of setting up inventories.

**Table 3.**
**Benefits Identified as per Stakeholder Impact Analysis [4]**

| Benefits | # |
|---|---|
| Opportunity to manage income & outgoings | 3 |
| Opportunity to offer new products/services | 2 |
| Improved status | 2 |
| Removal of tedious work | 2 |
| Improve satisfaction of work | 1 |
| Opportunity to develop new skills | 1 |
| Opportunity for organizational growth | 1 |

Following are the risks associated with migration to cloud. These risks can be summarized as Table 4 [4].

**Table 4.**
**Risks Identified as per Stakeholder Impact Analysis [4]**

| Risks | # |
|---|---|
| Deterioration of customer care & service quality | 3 |
| Increased dependence on external 3$^{rd}$ party | 3 |
| Decrease of satisfying work | 3 |
| Departmental downsizing | 2 |
| Uncertainty with new technology | 2 |
| Lack of supporting resources | 1 |
| Lack of understanding of the cloud | 1 |

Stakeholder impact analysis shows that although it is important to take financial benefits of cloud infrastructure it is equally important to consider the organizational dimensions a majority of management in organization are reluctant to use it beyond a certain environment as there are huge amount of risks associated. The outcomes from their case study show that cloud computing is a more cost effective solution than purchasing, installing and maintaining infrastructure. The results calculated by Khajeh-Hosseini et al.[4] show that the system infrastructure used in case study have a cost less than 37% in a span of 5 years using EC2 and support calls will be reduced by 21% of support calls for this system. These findings have a significant impact hence they can act as grounds for migrating to a cloud system but with the stakeholder impact analysis we can deduce that there is a significant amount of risk also associated with it. The benefits identified can be used to find out the gain buy-in from stakeholders and risks can be identified to find out whether the project is not getting affected negatively on migrating to cloud infrastructure.

## 4 CLOUD USERS

Cloud users play a very important role in any cloud computing application. They are the soul quality drivers [55]. The impact of any cloud solution depends on the impact it has on the cloud users. According to Vouk et al. [55] around four categories of cloud users are defined Cyber Infrastructure developers, Service authors, Service Integration and provisioning Experts and End Users also depicted by figure 3

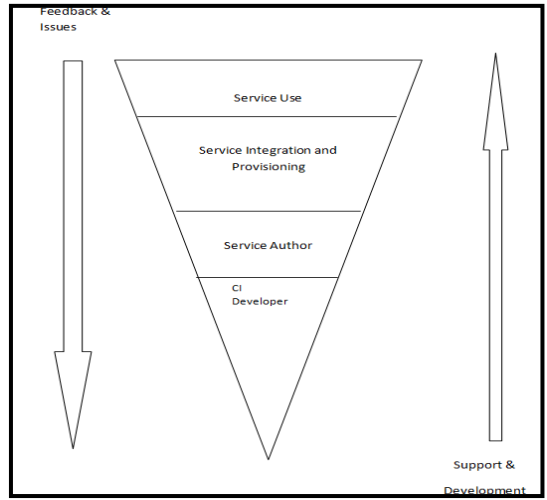

**Fig. 3. Cloud User hierarchy [55]**

1. **Cyber Infrastructure developers**: These are developers who are responsible for development, integration, administration, communication and maintenance of the cloud framework. They are specialists of their particular areas like networks, computational hardware, and storage. They are responsible for data abstraction in a cloud framework along with development of new functionalities for their cloud.
2. **Service authors**: They are developers who are involved in the task of developing base line images and services that can be integrated into workflows .These authors are the ones who are not an expert in cloud domain and therefore interfaces and tools available to them must be easy to learn and use.
3. **Service Integration and provisioning Experts**: These are experts who are responsible for providing complex solutions to its end-users. They use the services that are already existing and existing images are also used to develop newer solutions. They may also require some kind of expertise in the construction of images and services but their main focus is interfacing with end-users and provisioning them resources as per their requirements.
4. **End Users**: They are the most critical users of cloud services. Some of their requirements include provision of services in a reliable, easy to use and a highly available manner.

Experimental results have shown that increasing the number of users also increases the average response time of cloud services [85].This experiment shows that when the number of users is less than 500 than average response time is stable but it increases significantly when client size increases beyond 1000. PIQL proposed by Armbrust et al. [57] provides scale data independence i.e. it ensures that queries perform well both when the database size is less as well as when the database size increases. Table 5 shows a comparison between normal web hosting application users and cloud hosting users keeping in mind the special needs of cloud users.

**Table 5**
**A Comparison between Normal Web Hosting Application Users and Cloud Hosting**

| WEB USERS | CLOUD USERS | AUTHORS |
| --- | --- | --- |
| Fixed bill payment scheme by users | Bills are payed on a pay per use or metered manner by users | [106] |
| Users cannot allocate and reallocate resource i.e. users have no direct control over the use of their resources | Allows users to allocate and reallocate resources as per their requirement. | [106] |
| Users have access to a single server | Users have access to multiple | [106] |

|  | servers depending on their requirements at that point of time. |  |
|---|---|---|
| Storage space provided to such users is usually fixed or limited. | Cloud users usually have infinite storage space |  |
| Users perform management of data at their own premises. | Management of data is not in the hands of cloud users but lies with the cloud service providers. | [107] |

## 5 REFERENCE ARCHITECTURES FOR CLOUD

The architectural requirements for cloud computing are classified as per the user requirements of cloud such as enterprise users or end users along with service providers. There are currently several different architecture models available such as Amazon EC2, Cloud Security Alliance, Cisco, DMTF, NIST, Windows Azure etc. [7]. Liu et al. [105] have identified a cloud reference architecture which defines the various actors in cloud computing which include cloud consumer, cloud auditor, cloud service providers, cloud carrier, and brokers of cloud application along with their respective activities and functions in a cloud environment. In [56] a cloud database management system architecture has been proposed, which describes a three layered architecture for data management in cloud .In [7] generic logical data models (LDMs) have been presented which are independent of the existing architecture's implementation (of NIST and DMTF). These data models act as framework for requirement gathering in cloud architectures.

## 6 CLOUD COMPUTING PLATFORMS

There are different clouds computing platforms available currently. Few Cloud Platforms and their key features have also been depicted by Table 6. Some of them as enlisted by Mollah et al. [63] are:

1. **AbiCloud Platform:** AbiCloud is a cloud computing platform that has been built by a Spain based company Abiquo. This platform is used for building, integrating and management of public and private cloud in homogeneous environments [AbiCloud][12]. This platform will enable its users to automatically build, deploy and manage storage systems, servers, network, virtual devices etc. Also shown by Figure 4

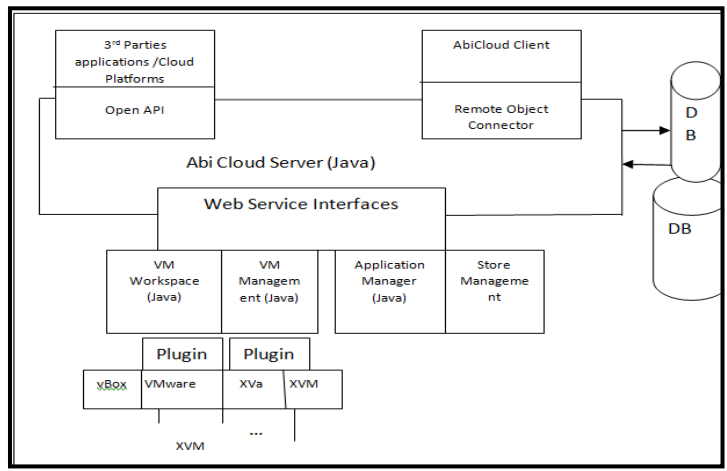

**Fig. 4. AbiCloud Platform [63]**

2. **Eucalyptus Platform:** Eucalyptus Platform has its roots from California University Santa Barbara [Eucalyptus][37]. It is an open source implementation of Amazon EC2.It is an Elastic open source infrastructure and uses cluster implementations of utility, cloud computing. It has become famous as a computing standard based on service level protocol. Also shown in Figure 5.

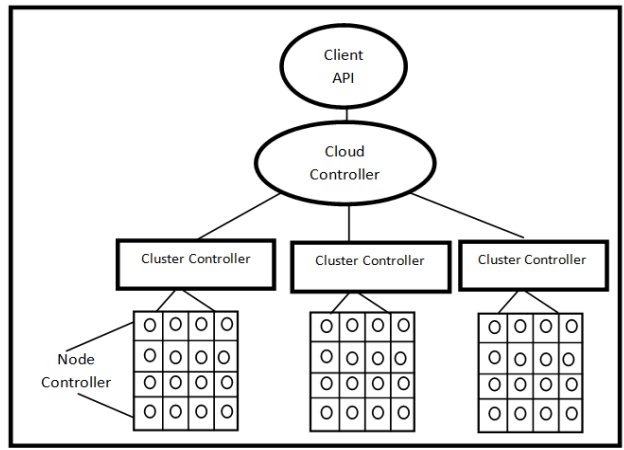

**Fig. 5. Structure of Eucalyptus Cloud [63]**

3. **Nimbus Platform**: It is an integrated set of tools that delivers the versatility of infrastructure as a service to scientific user. Its aim is to make moving to a cloud task as effortless as possible. It also aims at providing a bridge to users to act as stepping stone for building other resources like virtual clusters through resources provisioned by cloud [69]. Context Client, Cloud Client, reference client module and EC2 client modules are some of the client supported modules. Also shown in Figure 6

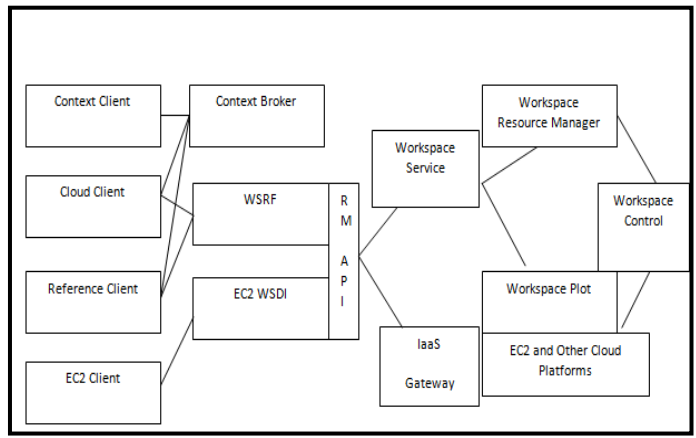

**Fig. 6. Structure of Nimbus Cloud Platform [63]**

4. **Aneka Platform:** Aneka is a cloud computing platform developed by Manjrasoft pty Ltd. It provides a framework for developing distributed applications that can be on cloud. Some of its key features include ability to provide different ways of expressing distributed application through its rapid deployment tools and framework, persistence, dynamic scalability, support for multiprogramming environment and security [ 65].

5. **Open Nebula Platform:** Open Nebula is an open source cloud service framework [62]. It aims at providing the most advanced and a highly scalable solution for building and managing virtualized data centers and IaaS Clouds. It also automates and orchestrates the operation of virtualized data centers along with ensuring stability and quality of software distribution [71]. Also shown in Figure 7

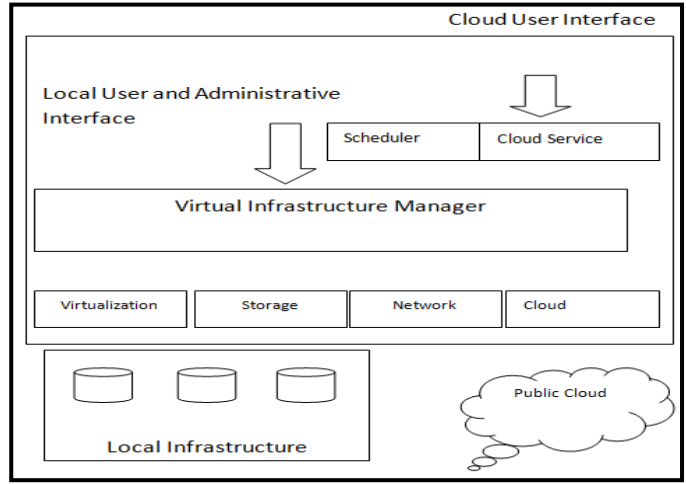

Fig. 7. Structure of Open Nebula Platform [63]

Table 6
Cloud Platforms and Their Key Features

| Platform | Programming Framework | Key Features | Services | Advantages | Authors |
|---|---|---|---|---|---|
| AbiCloud | Java, My SQL, Apache Tomcat, Network attached Servers | Suitable for building, integrating and management of public and private cloud, deep integration, powerful API, back office integration | IaaS | Provides its users with tools which can provide easy management, dynamic scaling of servers, and provisioning and re provisioning of servers. It also helps in saving costs of setting up datacenters. | [12],[113] |
| Eucalyptus | Java, Hibernate, Axis2C and Axis2 | Based on service level protocol | IaaS | Provides reduction in test costs, increased agility and availability , simplified process of performing software updates and improved image migration | [37] |
| Nimbus | Python and Jva | Provides infrastructure as a service to its users, support for best effort allocation, batch scheduling | IaaS | Specially designed for scientific community, provides user friendly concepts such as allowing virtual clusters on cloud | [69],[111] |

| | | | | provisioned resources. | |
|---|---|---|---|---|---|
| Aneka | C# and .Net supported API's | Rapid deployment tools, provides support for multiple runtime environments simultaneously | PaaS | Dynamic scalability, provisioning of resources based on QoS or SLA, optimization of capital expenses | [65],[110] |
| Open Nebula | Java, Ruby | Automatic orchestration of virtualized data centers, provides powerful schedulers for activities like load aware and packing | IaaS | Easily integrates with a wide variety of billing systems, provides hybrid cloud computing support through AWS connectors, Completely platform independent | [62],[71] |
| Amazon EC2 | Microsoft and Linux based, map Reduce | Persistent storage through simple storage services(S3), Amazon cloudwatch, elastic IP addresses, automated scale, uses Xen virtualization | IaaS | Pay per use(hourly), dedicated IP addresses, good bandwidth, no hardware failures | [110] |
| Google App Engine | Python, Java, PHP and Go | Google cloud SQL, Modules, Map Reduce Sockets, Google Cloud Storage Client Library | PaaS | Easy startup, automatic scalability, high security | [110],[111],[112] |
| Microsoft Azure | ASP .Net, Node.js and PHP | Can run windows as well as Linux systems | PaaS, IaaS | Hosted cloud services including high density hosting of websites, high availability | [110] |
| Force.com | Apex Language(Database service), C#, .Net | Database is handled through fields of relationships | SaaS | Provides seamless integration with other applications, has multilayered security features | [111] |

# 7 SECURITY IN CLOUD COMPUTING

According to NIST [70] cloud computing involves virtual environment which exposes the cloud data to several vulnerabilities and threats to users data privacy and security. Cloud computing though provides huge advantages but it also imposes a great amount of threat to security of data which is now stored to an off premise rather than an on-premise [19]. There are a variety of attacks that can occur on cloud computing environment, Some of them are [24] virtualization attacks such as VM escape and rootkit in hypervisor, man in middle attack, zombie attack phishing attacks and others. Some of the Top threats identified in cloud computing [27] [118] are:

- Data Loss/leakage
- Insecure API's
- Malicious Insiders
- Traffic Hijacking
- Abuse of cloud computing
- Unknown Risk profile
- Shared Technology vulnerabilities
- Distributed denial of services
- HTTP or XML based denial of service attack

Apart from these threats, lack of transparency between the cloud provider and clients can also be considered as a road block for people to move comfortably to the cloud [83]. In [116], a threat detection model has been proposed based on three goals detection of attacks; alert the parties and identification of type of attacks. Another security concern is data mining security attack in which users data is analyzed for a long period of time and then this data is used for extracting information about users thereby threatening their privacy [45].Since a single method cannot fully eradicate the problem of security in cloud computing therefore many new strategies are required to ensure security in a cloud environment [96]. A study of various security threats has been done by Shaikh and Haider [40] and some of the most vulnerable threats have also been identified. Thus security is a major concern which is leveraged on cloud service providers and eradicating these threats to security is a matter of prime importance for cloud service providers in order to attract more and more probable clients. SACS, a security model based on hadoop map reduce framework is more stable in case of a security threat [39]. Four indicators of vulnerabilities that are specific to cloud have been presented by Grobauer et al. [16].Currently various challenges to cloud security are available [25]. An SLA and accountability can be considered as the building blocks for security of data in cloud. In [117] a rusted third party cryptography based solution is proposed which exploits the use of public key .Leakage prevention solutions in cloud can also be considered as an effective security framework. Multitenancy trusted computing environment model (MTCEM) is a model ensuring trusted cloud infrastructure to its customers developed on IAAS platform by Li et al. [98].Advanced Cloud protection system proposed by Lombardi and Pietro [119] guarantees of an enhanced security against most of the existing known attacks and has been tested on eucalyptus and OpenECP cloud platforms.

## 7.1 Monitoring Intrusion and Security breaches

One of the major challenges which have been identified in relevance to cloud computing is the security challenge. There is a lack [11] of clarity of perimeters, confidentiality, ownership and integrity of data in cloud. As per a survey conducted by Defcon [35] amongst 100 IT professional at 2010 DEFCON conference 96 percent of the IT professionals claimed that cloud will provide an increase in hacking activities as the cloud providers still fail to provide a secure cloud infrastructure. Therefore, it is the need of the hour to define and utilize a threat monitoring mechanism for cloud services to deal with intrusion is to cloud systems[121] such as insider attack, port scanning, flooding attacks and attacks on the hypervisor itself. Some of the securities monitoring mechanisms available in the market today are [35]:

1. **Commercial Solutions**: different companies adopt different security monitoring mechanisms for example Amazon uses cloud watch (a web service that monitors cloud components and retrieval of statistical data).
2. **Vulnerability reporting process** (when vulnerability is found in AWS products). Carl et al. [42] have presented a denial of service survey which provides detailed information about detecting DoS flooding attacks. Various attacks can be Distributed DoS attack in this the assault is carried out at different hijacked system by a single attacker, Network based DoS attacks etc.

Open communities are communities which have come up with a solution for open standards which can be used in cloud infrastructure. A few of these are
- Cloud audit/A6:It enables cloud service provider to automate and audit assurance of their service models for authorized users through their namespaces and interfaces [22].
- Cloud security alliance: It is a nonprofit organization engaged in the task of developing security tasks in cloud computing.
- Open cloud computing interface working group (OCCI-WG): The task of provisioning and monitoring of cloud infrastructure is carried out by this group along with taking into account factors like interoperability, portability, integration.

Since cloud computing defines newer challenges than the existing technologies therefore it is required to modify the existing monitoring mechanisms [11]. Cloud support teams should have access to all the components of cloud infrastructure in order to provide solutions. Elasticity and scaling up and down of resources is not well supported with the existing mechanisms. Cloud computing has to handle data from different customers therefore mechanisms for monitoring in such an infrastructure is also governed by the security concerns of individual clients. Cross layer monitoring [11] can also be treated as an effective monitoring mechanism with several advantages like avoiding duplication of tasks in varied layers, more accurate monitoring. A PMU based state application suggested by Maheshwari et al. [51] can be used for mapping targeted system to cloud architecture.

## 8 CLOUD FOR STORAGE OF DATA

Storage of data in cloud enables users with a facility to access data as per their requirements irrespective of their time and location. In cloud, data is stored at data centers in form of clusters of raw data servers which enables retrieval of this raw data [104]. Cloud users find it very convenient to move their data to cloud because they no longer have to worry about making any huge investments for hardware infrastructure and their maintenance and deployments [78] .Storage services in cloud involves delivery of data storage as a service. The resources are often provided through a utility computing like basis i.e. the part of the cloud resources a user is using is the only part he is charged for, it can be for a month, a week or just a day. In[114] an optimal cloud storage management system is introduced. The biggest advantage of cloud storage is the geographical independence that it offers i.e. the data in cloud can be accessed from any location at any time using any computing device. For example users can pay the cloud service provider for storing some of his important data that he/she might need to retrieve at a later instance of time at some other location using some other device. Since the data in cloud is stored at some external location away from the users therefore this data is prone to several attacks by external sources like some unauthorized person or by some internal sources which can be due to some untrustworthy service provider etc. PosixCloud developed by Xu et al. [76] is a storage cloud with enhanced metadata scalability. In order to ensure proper hosting of data in cloud an efficient data storage auditing protocol for cloud is introduced by Yang and Jia [52]. Cloud Zone is a cloud data storage architecture based on multi agent system architecture and consists of two layers cloud resource layer and MAS layer [9].Several protocols are present to ensure access to data present in cloud. A survey of these protocols has been analyzed by Priyadharshini and Parvathi [17].From [30] we can get an overview of state of art for various cloud storage approaches. In a cloud setup the data which is available is huge and in unstructured format this kind of data is difficult to be stored and managed by fixed and structured data models which are made available by Relational database Management System. This has led to the development of No SQL system or Key Value Store systems and will provide the advantage of simplifying roll outs. Some of the NO SQL database systems as discussed by Chandra et al. [33] are:
1. **Cassandra**: Apache Cassandra database is preferred in case where high scalability and high availability is desirable. It uses a column family information model based on column indexes and provides support for log structured updates and built in caching. In order to handle huge and interactive datasets companies like Netflix, Twitter, Urban Airship, Reddit, Cisco use Cassandra.
2. **HBase**: It is an open source, distributed and column oriented model based on Google's big table. It is well suited for applications that require random and real time access read/write access to big data.
3. **Mongo DB**: It is a highly scalable, high performance, open source and NOSQL database written in C++.It is fault tolerant, persistent, schema free and document oriented implementation of Mongo DB.

4. **PNUTS**: It is a large scale, hosted and centrally managed database system .It is based on data serving for web applications like offline analysis of web crawlers. It reduces application development time and provides data management as service.
5. **Big Table**: Big table maps two arbitrary string values. It can easily scale up and add huge number of machines as per the requirements without any reconfigurations.
6. **Windows Azure**: Windows Azure is an open cloud platform that enables users to instantly develop, deploy and manage applications over Microsoft data center.

Apart from these NOSQL database systems HSQL (Highly scalable Cloud) by Chang et al. [20] is a database which uses a distributed B-tree column indexing scheme is a scheme for supporting indexing for non row key columns along with parallel B-tree search for HBase, in huge database. HSQL enhances the existing HBase by providing an SQL query interface, along with a distributed B-tree indexing and support for transactional data processing. The experimental results have shown that the response time achieved by this indexing method is better than an HBase even with large data and large number of queries. In comparison to HBase scan filtering has an average of 6.2X speed up for range query and an average of 12.2X speed up for aggregation. Cloud services on the basis of data flow can be categorized into two main categories Transborder cloud services and domestic cloud services [31].In domestic clouds the entire cloud is physically located under the same jurisdiction whereas in a trans border one involves one or more jurisdictions.

## 8.1 Shared Nothing and Shared disk Cloud database architectures

In cloud computing storage plays a pivotal role in data center and cloud services. In cloud computing storage virtualization allows storage to be allocated and reallocated dynamically as per the requirements of the users. There is still no well defined architecture for cloud computing till date. There are two primary architectures that can be used in cloud computing shared nothing and shared disk architectures. Cloud computing allows abstraction of complex data storage from the cloud consumers. This abstraction of cloud data has performance concerns in case of a cloud computing environment where there are multiple tenants distributed across several geographical locations and there data is also distributed across several servers. Shared disk and shared nothing are data access architectures [81].In shared nothing environment each user has its own private memory and disks. Shared Nothing partitions the data such that each server is in charge of maintaining its part of database. It allows ownership of data to be dynamically transferred to another system. Shared nothing approach for implementing cloud database via approach is advantageous because it provides dynamic scalability but it lacks the feature of dynamic load balancing i.e. each server has to endure peak load for its data. In shared disk any number of database nodes can process any part of data. In this model all the system processes have access to all the resources of system and the data i.e. all the systems connected share the same disk. One of the demerits of this approach is that it does not scale as well as the shared nothing architecture. It is well suited for applications that require only a little amount of data sharing or for workloads that are very difficult to partition. Since shared disk allows access to the entire database, hence any data request can be handled by any node thereby enabling a fluid load balancing. It also allows you to run each server at a high CPU utilization.

It should be taken into consideration that shared disk architecture requires less number of low cost servers and has comparatively less maintenance cost and can be considered suitable for cloud environment [93]. But, both shared disk and shared nothing have some disadvantages along with certain advantages and it is up to the enterprise to choose a particular architecture depending upon its requirements from its cloud database.

### 8.1.1 Comparison of scalability, high availability, loads balancing and data consistencies
1. **Scalability:** is defined as the ability for a system to increase and decrease its workload dynamically as per the user requirements. Shared nothing partitions is better suited for applications that require dynamic scalability but partitioning scheme ,function and data shipping volume can become a bottleneck. Shared disk also has scaling challenges driven by inter nodal message passing. The more nodes that are added the longer the wait states.
2. **High availability**: support those applications that can provide continuous service even in case of failures. Shared nothing approach consists of master nodes supported by more than one slaves per master. These slaves become masters on failure of a master node. Shared disk can respond to failures by routing database requests to next available node. The server failure does not hinder the performance but instead balances the load on the remaining servers connected to system.
3. **Load balancing**: is a technique that allows sharing of load or work between several computers, processors or hard drives. Shared disk does not support dynamic load balancing whereas shared disk allows fluid load balancing which enables it to make any amount of changes in usage patterns.
4. **Data consistency**: allows information accessed on a network to be valid at all times .Shared disk ensures data consistency at all times as the data is shared amongst several processes whereas shared nothing requires some efforts like configuring the master server to reduce read performance and synchronization of all the transactions.

It should be taken into consideration that shared disk architecture requires less number of low cost servers and has comparatively less maintenance cost and can be considered suitable for cloud environment [93]. But, both shared disk and shared nothing have some disadvantages along with certain advantages and it is up to the enterprise to choose a particular architecture depending upon its requirements from its cloud database.

## 8.2 Cloud Data base management system

DBMS available on cloud can be of varied forms apart from relational. For example Google's big table is not relational; Microsoft SQL Azure is a fully relational DBMS. Cloud Data base management system is a distributed database that enables computing resources to be made available as service via an internet connection rather than as a product. In cloud dbms applications are connected to a database which is available on cloud. Native cloud data bases are considered to be more efficient, available, elastic and scalable than the traditional ones. SQLMR [58] is a scalable data management system for cloud .SQLMR compiles SQL to its Map Reduce counterpart. It is also compatible with existing SQL based queries. In [Gelogo and Lee][102] a DBMS in cloud architecture is proposed, It identifies three layers in cloud dbms .Storage is the first layer, database form a layer above it and the topmost layer is the application layer and consists of web servers, application servers, access management etc. It provides well distribution of data along with efficient data access. Data is stored in an encrypted form during storage and hence no programming is required for its encryption or decryption.

Some of the key responsibilities at the end of cloud service provider of data are [15]
1. **Scalability**: cloud database providers must be able to support very large database, with low latency rate. Vaquero et al. [54] provides a survey about application scalability in terms of platform scalability, server scalability, network scalability and database scalability.
2. **Data replication**: issues regarding data replication are to be handled by the cloud service providers.
3. **Recovering from failure**: in case of any transactional failures or any other failures due to networks etc. cloud service provider is responsible for detection of such a failure and is also responsible for rectifying it.
4. **Allocation or reallocations of servers**: Its the responsibility of the cloud service provider to increase or decrease the number of allocated servers as per the users requirements.
5. **Availability**: cloud database must be made available at all the times irrespective of any failures due to network, or non availability of data centers i.e. cloud services must have high failure tolerance levels.
6. **Privacy**: privacy of the user's data must be ensured at all times by the service providers.
7. **Securing data**: cloud clients needs to just enter their data and after the client has entered his or her data; it's then the responsibility of database provider to ensure security of the data of its client.
8. **Metering**: cloud service providers must be able to properly estimate the amount of usage a particular client.
9. **Service availability across geographical locations**: The cloud services must be made available across the world irrespective of the geographical locations.
10. **Simple APIs**: Simple and user friendly interfaces must be made available to the clients.
11. **Operational ease**: Cloud based systems must be easy to operate, and must provide the customers with a user friendly interface.

In cloud computing multiple distributed data stores are supported automatically but it has certain tradeoffs which can be provided by a database support layer [18]. ES2 [100] is an elastic data storage system on cloud platform which supports both analytical processing as well as transactional processing. In order to ensure security of data stored in cloud and to assure correctness of data a two way handshake method based on token management has been proposed by Tribhuwan et al.[64]. Apart from insuring storage correctness this method also performs data error localization.

### 8.3 Oracle Database 12c

Oracle 12c is a database developed specially for cloud environment based on multitenant architecture. It enables the users to treat several databases as a single database but preserves isolation and resource control amongst the databases. It uses heat maps to identify access patterns.

It provides several advantages and features which are inevitable for a cloud environment. These features include the following [72]:
1. Usage pattern based optimized automatic data storage and data compression.
2. Provision for multitenant containers.
3. High availability and accessibility through Oracles maximum availability architectures.
4. Reduced IT costs and complexity.
5. Improved IT services
6. Support for big data.
7. Security of data.
8. Support for resource sharing.
9. Better administration as fewer databases need to be maintained by the administrators.
10. Rapid database provisioning and cloning.

### 8.4 Database as a Service

In Database as a service, the database is held within a cloud environment and is accessed on a pay per use mechanism as a services via an internet [92] .Relational cloud is a transaction based database as a service scheme [21]. It has the following characteristics (1) provides division of workload on a database server thereby achieving better performance, and (2) agile security criteria's that can run SQL queries over encrypted data .Cloud data on the basis of cryptographic parameters as per the clients specifications can be classified as Confidentiality, availability and integrity [86].This division of cloud data provides better protection and access of data..

Many cloud computing providers such as Amazon RDS, Microsoft SQL Azure, salesforce.com, database.com, Xeround etc. are now providing Database-as-a-Service. It provides features such as multitenancy, pay per use and elastic resource allocation and reallocation. Though it provides several advantages to the users such as reduction in operational costs, elasticity and on demand database requests etc., It has a few draw backs like it is very difficult to predict the kind of database requests a user might come up with and it is even more challenging for the database as a service providers to predict the database requests as well as when and where to place database in cloud depending on the resource requirements. Apart from these drawbacks maintaining consistency amongst different replicas of the same servers is also a challenge which can be handled by a tree based consistency approach [61]. According to the database as a service providers view the main goal of database as a service is to achieve an efficient resource allocation among multiple databases and to minimize the problems that are caused due to migration of an enterprises database to cloud and to maximize resource utilization in cloud. Yu et al. 2012 [89] have given a cost efficient algorithm for placement of resources in an efficient manner has been proposed taking into consideration multi resource constraint, user and system preferences. Its main objective is to help the DBaaS providers to achieve an efficient resource allocation amongst multiple databases, provide utilization of cloud resources in an efficient manner and minimize problems that erupt from migration of a company's database to cloud.

Cloud databases can be very efficient for holding databases and can also be an effective mechanism for holding query services as it provides the end users with dynamic scalability along with reduced costs. In [46] RASP an efficient data distribution method is proposed which provides a secure and efficient query processing scheme which can significantly reduce the workload and uses methods like encryption via order preserving, random projection etc to safeguard distributed queries and data exposed in a cloud environment. Smart SLA is a cost aware resource management system for database in cloud environment [75] and it has two major components system modeling module (based on machine learning) and resource allocation module(adjusts resources in a dynamic manner).

Return on investment can be used as an effective tool by business executives for assessing purchase decisions for cloud DBaaS set up [92]. It enables the organizations to decide upon things like whether their database is suitable for moving towards cloud infrastructure. In fact now, many leading companies have their own ROI and TCO calculators available online. Microsoft's [66] online ROI and TCO calculates ROI and TCO for deploying an organizations database on Widows Azure Platform in a customized manner as per the client requirement's like currency, country etc. Amazon [13] also has an online calculator for calculating the costs incurred by users for using various amazon web services like Amazon EC2, Amazon RDS, and Amazon Simple DB etc.

### 8.5 Limitations of Data management in cloud

Though cloud computing is an emerging technology with ample of use case and advantages but its not free from limitations .Some of its limitations are:
1. **Lack of Privacy** [28]: Since data is exposed by users or organizations to a third party host which is outsider to organization .Therefore privacy is no longer held.
2. **Connectivity and mobility** [28]: Since cloud computing requires network connectivity between users and service provider's .This connectivity is not always available at places where there are no networks such as airplanes etc.
3. **Data is stored at untrusted host** i.e. a third party [34]: Since the cloud service provider is a third party organization therefore it's not trust worthy at all times as it can misuse the information being hosted in its cloud.
4. **Lack of user control** [28]: Since data is stored at third party host, therefore users have no control over their data once it's in cloud.
5. **Data is duplicated** across several geographical locations thus, making it more prone to malicious attacks [34].

## 9 EMERGING CLOUD COMPUTING FIELDS AND TERMINOLOGIES

With the increases in popularity of cloud computing several newer terms and fields have now emerged in cloud computing. Some of these are

### 9.1 Green Cloud Computing

Since cloud computing is a rapidly expanding technology which is spreading across the globe nowadays therefore as it grows the energy consumption of network and computing resources is also growing at a fast pace. Majority of this energy is consumed in switching, transmission, and processing and data storage. Cloud computing can be used for reducing the energy consumption where tasks are low intensity ones [48].Green cloud computing is now an emerging field .Objective of Green cloud computing is to provide an elastic, flexible, oriented framework along with reduced energy consumption[50]. Recently works done on green cloud computing are based on networks include cooling systems, virtualization, task scheduling algorithms etc [68]. Chu et al. [41] have given an analysis of cloud computing energy consumption on the basis of type of services. They have also suggested that green mobile communication is the pillar on which green cloud computing will stand. An energy aware layer which can evaluate energy consumptions at data centers in micro and macro metrics and then migrate services to host consumes energy in cloud computing more efficiently [79]. Architectural strategies for green cloud computing are discussed in [74]. A dynamic environment for provisioning of resources in a virtualized can achieve power efficiency by using a workload predicting and optimum power efficient allocation [80]. Green Power management given by Yang et al. [26] provides load balancing for virtual machines in cloud environment. An approach for a less energy consuming cloud computing data center for developing regions is proposed by Baikie and Hosman [14].

### 9.2 Cloud bursting

Cloud bursting is a new use of cloud resources. As per Bicer et al. [88] Using this concept an organization can meet up to its daily resource requirements using its own resources as well as use resources provided by cloud service providers during peak hours to meet up to its resource requirements. Cloud bursting can also safeguard organizations from over provisioning of resources thereby providing its users with a better response time. Challenge of data intensive computing with cloud bursting has been used by Bicer et al. [88] for designing of a software framework that allows data intensive computing through cloud bursting, in this a combination of available resources are used along with resources from the cloud to perform map reduce type processing on geographically dispersed and distributed data. He also describes a middleware that supports map reduce type API in a cloud bursting environment. This middleware can also support transparent remote data analysis paradigm. Through this approach overhead involved in inter cluster communication is low for majority of data intensive applications. The load of computation is also well balanced by this middleware. Therefore, cloud bursting provides a flexible approach for combining different base resources with cloud resources along with taking the advantage of pay per use mechanism offered by cloud applications; it can easily scale up and down of cloud applications to provide a secure cloud environment. One of the major security concerns for cloud computing is to reduce distributed denial of services [73]. Cloud Bursting Brokerage and Aggregation (CBBA) [74] describe an approach proposed for a multi cloud environment based on concept of classes and objects. In this approach four phases have been defined.

**Phase 1**: Cloud 1(C++ Cloud): In phase 1 aggregation and bursting is done on C++ files according to values based on object oriented properties such as class and objects.

**Phase 2**: Cloud 2(Java Cloud): In phase 2 aggregation and bursting is done on java files according to values based on object oriented properties such as class and objects.

**Phase 3**: Cloud 3(C# Cloud): In phase 3 aggregation and bursting is done on C# files according to values based on object oriented properties such as class and objects.

**Phase 4**: Sharable cloud: In this phase sharing between clouds with security keys is done, such that only authorized cloud users can access and share resources from one cloud to another so that resources of one cloud is utilized in another.

### 9.3 Cloud Federation

In Cloud federation the service providers make a mutual agreement about policies and governance by establishing a common trust boundary [115]. A federation of cloud service providers consisting of SaaS, PaaS and IaaS can leverage multiple independent clouds [32].This federation provides enhanced consumer values. In [32] a weather forecasting technique is used to exemplify the concept of federation and delegation. Cross cloud federation manger can overcome several limitations of cloud computing. It is a solution which allows clouds to be federated with other types of clouds using a three phase model which involves steps such as discovery, match making and authentication [3]. Data web can also support cloud federation [2] and addresses issues like logical mapping of virtual resources and how to allow only authorized resources to access data. Li et al. [95] describes an SDO architecture that contains the general sequence in which service deployment takes place, for example service construction, IP discovery and filtering, negotiation, IP assessment, Service contextualization, service data upload, SLA creation and service resource update etc.

### 9.4 Mobile Cloud Computing

Mobile cloud computing is a cloud computing area that has gained significant importance from the research community, but it is still at a nascent stage [126]. Mobile cloud computing is defined as a blend of mobile networks and cloud computing. It takes the advantages of elasticity of resources, storage, heterogeneity and pay per use model of cloud computing for numerous mobile devices across the globe [127]. It opens a world of new opportunities for the mobile network operators

## 10 RESEARCH CHALLENGES IDENTIFIED

This section highlights the several open issues in cloud computing. There are various challenges that have been identified after review of the related literature. These are the issues that have not been solved completely. These gaps can provide useful directions for future work:

1. **Data Management**: The data in distributed systems is increasing at an exponential, this led to the term big data [122]. Since cloud computing is now being directly linked with big data , the current requirement is to handle such kind of data via cloud by taking advantages of several useful features of cloud such as scalability and elasticity. These require tasks such as distribution of huge volumes of data across distributed data centers, monitoring the storage pattern of data, support for distributed transactions though CDBMS has been introduced on its lines. There is also a need to validate the data which is being integrated for storage .apart from these there are security and privacy issues which hinders effective and reliable data management and storage in cloud.   Data management challenges also include isolation of data's to ensure security of data

   stored and authentication of data access.
2. **Management of resources by the service providers:** Since resource requirement in cloud varies from time to time [122] .The main concern for any cloud service provider is optimal utilization of resource. This deal with energy efficient management of resources, future prediction of resource requirement's based on the access pattern, provision for elastic scale up and scale down of resources. Apart from these requirements there is also a need for insuring that the availability of resources is fault tolerant against peak resource requirements.
3. **Cost Estimation Model**: One of the reasons why cloud computing has evolved and has gained popularity is due to the pay per use model where computing resources can be provided like a commodity. Therefore we must address the need for development of effective cost models which can take care of all the billing, monitoring, accounting and auditing. This deals with transparency between delivery of services and billing, monitoring of SLA's and data processing.
4. **Development of Software Frameworks:** Cloud computing deals with data intensive and distributed applications, most of the applications at present make use of frameworks based on Map reduce such as Hadoop which provides a highly scalable and fault tolerant processing of data [123], but the main drawback of using hadoop is that it is highly dependent on the type of application being used. Some of its tasks may require intensive I/O operations while others may involve large number of CPU resources. Furthermore the VM's used in each of the nodes in hadoop are heterogeneous in nature. Therefore there is scope for improvement in such a scenario by carefully selecting the configuration parameters in such a system. Besides these there are other challenges in such frameworks like performance modeling, design of efficient algorithms along with making these frameworks energy aware.
5. **Decentralization of management systems for VM's**: the management system of virtual machines should be decentralized and must facilitate deployment of thousands of machines concurrently [124]. The cloud migration mechanisms must also be decentralized and the system must be robust and reliable at all times.
6. **Reduction in energy consumption:** this is another research challenge which has gained significant momentum. Cloud computing involves the use of several machines and other resources therefore there is a need for reduced energy footprints. Thus, we require development of appropriate metrics for energy consumptions for proper SLA management along with management of energy at all levels which include virtual machines, federations and product site [124].
7. **Service Provisioning:** Service provisioning deals with service availability which is robust, development of service provisioning architectures which are scalable in nature, efficient and flexible. It also deals with development of resource allocation algorithms which deals with optimal allocation of resources.
8. **Interoperability:** Interoperability is an issue that hinders the smooth transition of data between the different clouds or between a cloud vendor and user. Interoperability is a prerequisite at different levels in cloud. This has lead to a phenomenon where the usage of cloud services is not clear and depends on the cloud offering being used referred to as "Hazy Cloud" [125].

### 10.1   Proposed Architecture

We now propose a cloud architecture shown in **Fig. 8.**, which ensures privacy to cloud users and provides a low cost and secure cloud environment. The proposed system mainly consists of five modules, (i)Privacy and security management module, (ii)Resource allocation and load balancing module, (iii)scheduling module,(iii) cost estimation and negotiation module and (iv)data storage and management module. The privacy and security management module is responsible for ensuring user privacy and security of cloud users, their data and their transactions. This module also determines the most appropriate policies that need to be enforced. For instance if a cloud setup is being used by a group then authentication and access settings need to be changed. Resource allocation and load balancing module is responsible for allocation of resources to the cloud users, monitoring nodes that have potential to handle varied clients and also for load balancing and distribution of load at different nodes. The scheduling module consists of job scheduling managers that are in charge of maintaining pool of jobs, partitioning of jobs amongst the different nodes etc. The cost estimation module is responsible for calculating the costs associated with the usage of resources depending upon the amount being used, the priorities and privileges associated with a particular user in a transparent manner such as execution speed and high priorities. The data management module is responsible for management and storage of data at the data centers. Each of the modules should be interoperable with each other lead to a dynamic ,secure and more manageable cloud environment.

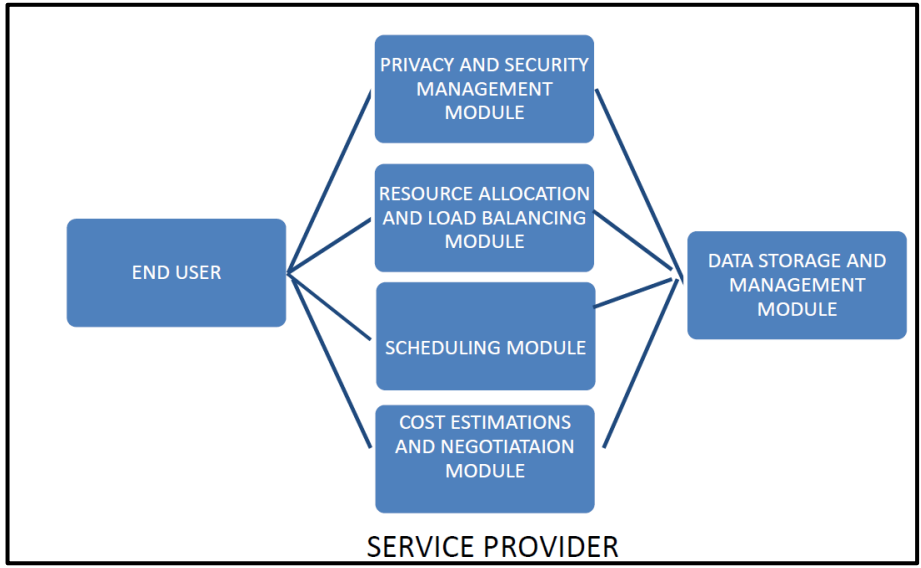

**Fig. 8. Proposed Architecture**

## 11 CONCLUSION AND FUTURE

Cloud computing is use of computing resources as a service via an internet. It is a very promising technology for the future with several advantages like pay per use, availability, elasticity etc. In this paper we have discussed the various developments that have taken place in this field. There are four different models for deployment of a cloud: public, private, hybrid and community. Cloud Service providers offer their services through several service delivery models. The various cloud service models are: Software as a Service, Platform as a service, database as a service and Infrastructure as a Service. Merits and demerits of migrating applications and infrastructure of an organization must be considered before a realistic migration process is carried out. The most pivotal role in any cloud environment is played by the cloud users. The impact of any cloud solution depends on the impact it has on the cloud users. There are four categories of cloud users are defined Cyber Infrastructure developers, Service authors, Service Integration and provisioning Experts and End Users. There are numerous cloud computing platforms available at present. Some of them are Abi Cloud, Eucalyptus, Nimbus, and Open Nebula. Cloud bursting is a new use of cloud resources .In this organization uses its own resources to meet its daily needs but it can also use resources from cloud service providers during peak hours to meet up to its resource needs. Monitoring intrusion and assessing security breaches is a critical task to overcome threats to security in cloud. Cloud users find it very convenient to move their data to cloud because they no longer have to worry about making any huge investments' for hardware infrastructure and their maintenance and deployments. Traditional database management systems are not able to handle cloud data and therefore NOSQL based data bases like Bigtable, Apache CouchDB, Windows Azure etc have been developed. Shared Nothing and Shared disk Cloud database architectures are the architectures which are used in cloud computing. Cloud Data base management system is a distributed database that enables computing resources to be made available as service via an internet connection rather than as a product. DBaaS is a flexible resource offering and pricing model. Thus cloud computing is an emerging technology which is here to stay. Though it has some limitations like security issues, high availability issues still it has huge scope and future. In future XaaS will be the service model which will be used which will provide everything as service. Also, security of cloud can be considered as a significant research problem that needs to be addressed.

## ACKNOWLEDGMENT

The authors would like to thank Mrs Seema Shakil for her continuous support during the course of work.

**Mansaf Alam** received his doctoral degree in computer Science from Jamia Millia Islamia New Delhi in the year 2009.  He is currently working as an Assistant. Professor in the Department of Computer Science, Jamia Millia Islamia. He is also the Editor-in-Chief, Journal of Applied Information Science. He is in editorial Board of some reputed Interntional Journals in Computer Sciences and has published about 24 research papers. He also has a book entitled as "Concepts of Multimedia, Book" to his credit. His areas of research include Cloud database management system (CDBMS), Object Oriented Database System (OODBMS), Genetic Programming, Bioinformatics, Image Processing, Information Retrieval and Data Mining.

**Kashish Ara Shakil** has received her Bachelor's degree in Computer Science from Delhi University in 2008 and also holds an MCA degree(2011) as well. She is currently pursuing her doctoral studies in Computer Science from Jamia Millia Islamia ( A Central University). She has written about three research papers in the field of Cloud computing. Her area of interest includes database management using cloud computing.